\pdfoutput=1

\documentclass[aps,prl,amsmath,amssymb,floatfix,twocolumn,amsmath,superscriptaddress,twocolumn,nofootinbib,tighten,letterpaper]{revtex4-2}

\usepackage[utf8]{inputenc}
\usepackage{bm}
\usepackage{siunitx}
\usepackage{comment}

\usepackage{enumerate}
\usepackage{amsfonts}
\usepackage{amsmath}
\usepackage{amssymb}
\usepackage{color}
\usepackage{soul}
\usepackage{float}
\usepackage{bm}
\usepackage{graphicx}
\usepackage[caption=false]{subfig}

\usepackage[colorlinks,allcolors=blue]{hyperref}

\let\phi=\varphi
\let\epsilon=\varepsilon

\renewcommand{\vec}[1]{\bm{#1}}

\definecolor{DarkRed}{rgb}{0.80,0,0}
\definecolor{DarkGray}{rgb}{0.7,0.7,0.7}

\newcommand{\prlsection}[1]{\textit{#1}.\kern0.05em---\kern0.05em\ignorespaces}
\usepackage[capitalise]{cleveref}

\begin{document}

\title{Emergent metallicity at the grain boundaries of higher-order topological insulators}

\author{Daniel J. Salib}
\affiliation{Department of Physics, Lehigh University, Bethlehem, Pennsylvania, 18015, USA}

\author{Vladimir Juri\v{c}i\'c}\thanks{Corresponding author:juricic@nordita.org}
\affiliation{Departamento de F\'isica, Universidad T\'ecnica Federico Santa Mar\'ia, Casilla 110, Valpara\'iso, Chile}
\affiliation{Nordita, KTH Royal Institute of Technology and Stockholm University, Hannes Alfvéns v\"ag 12, SE-106 91 Stockholm, Sweden}

\author{Bitan Roy}\thanks{Corresponding author:bitan.roy@lehigh.edu}
\affiliation{Department of Physics, Lehigh University, Bethlehem, Pennsylvania, 18015, USA}

\begin{abstract}
Topological lattice defects, such as dislocations and grain boundaries (GBs), are ubiquitously present in the bulk of quantum materials and externally tunable in metamaterials. In terms of robust modes, localized near the defect cores, they are instrumental in identifying topological crystals, featuring the hallmark band inversion at a finite momentum (translationally active type). Here we show that GB superlattices in both two-dimensional and three-dimensional translationally active higher-order topological insulators harbor a myriad of dispersive modes that are typically placed at finite energies, but always well-separated from the bulk states. However, when the Burgers vector of the constituting edge dislocations points toward the gapless corners or hinges, both second-order and third-order topological insulators accommodate self-organized emergent topological metals near the zero energy (half-filling) in the GB mini Brillouin zone. We discuss possible material platforms where our proposed scenarios can be realized through the band-structure and defect engineering.
\end{abstract}

\maketitle


\emph{Introduction}.~Topological lattice defects are ubiquitous in crystalline materials and are of central importance for their structural properties. In the last decade, they have emerged as viable platforms for probing topological phases of matter through a subtle interplay of the topology of the electronic wavefunctions and  lattice geometry. In particular, dislocations, the defects associated with lattice translations, can probe a wide range of topological crystals  featuring the band-inversion at a finite momentum directly in the bulk via the symmetry and topology protected localized defect modes either at zero~\cite{Ran-NatPhys2009, Teo-PRB2010, Juricic-PRL2012, Asahi-PRB2012, slager-natphys2013, HughesYao2014, Slager-PRB2014, Nag2021, das2022dynamic, PanigrahiPRB2022, Panigrahi2022, hu2022dislocation} or finite~\cite{Roy-Juricic-PRR2021} energy. As such, these modes are immune to interface contamination and surface termination, and have been experimentally observed in both topological crystals~\cite{Hamasaki-APL2017, Nayak-SciAdv2019} and their metamaterial analogues~\cite{Dislocation-Acoustic2021, Ye2022}.

\begin{figure}[t!]
\includegraphics[width=0.95\linewidth]{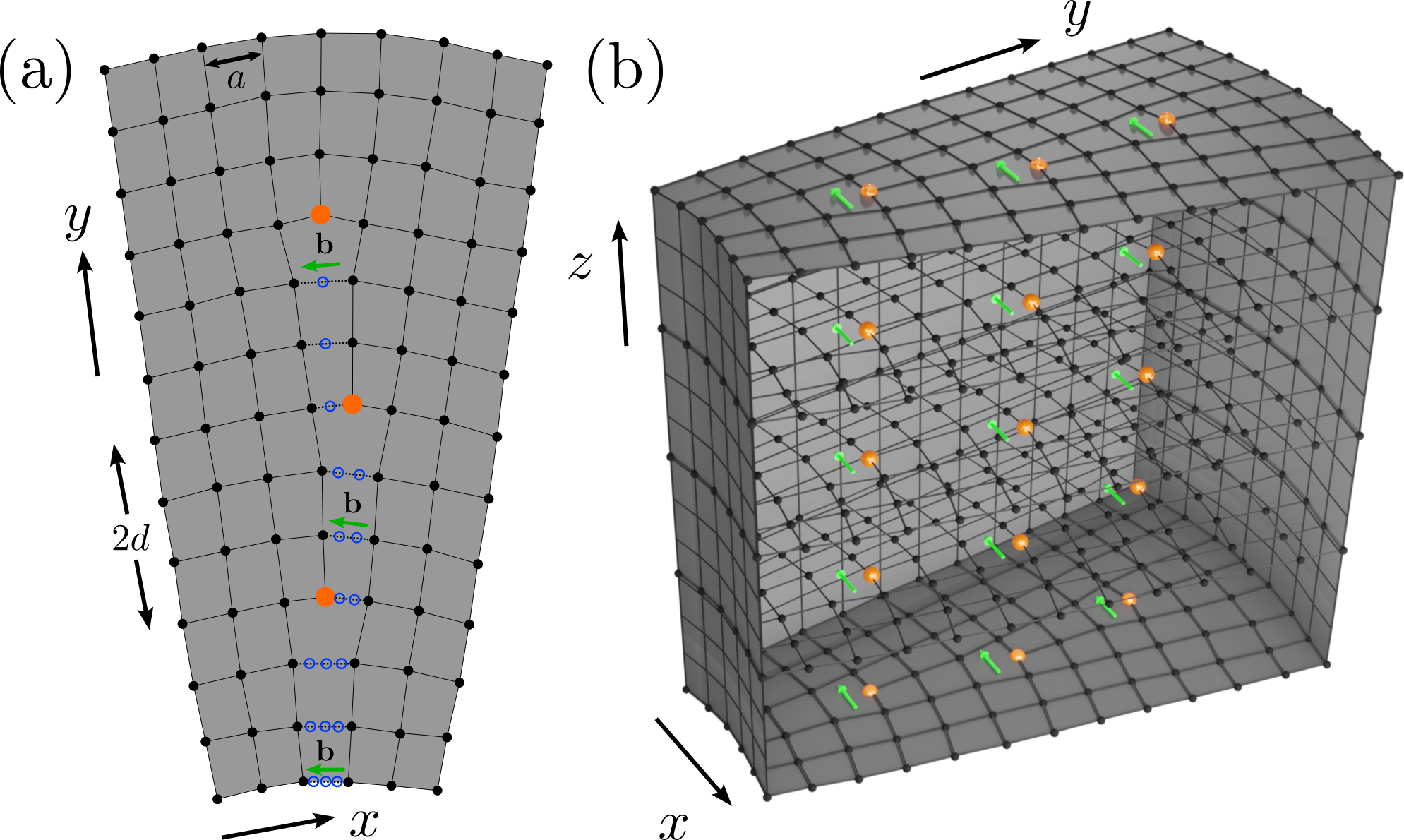}
\caption{Grain boundary of edge dislocations with Burgers vector ${\bf b}=a{\bf e}_x$ in (a) two and (b) three dimensions, where $a$ is the lattice spacing. The center of individual dislocation cores are shown in orange and the missing sites are shown by open blue circles in (a). The distance between successive dislocation core is $2d$. See text for additional details.}
\label{fig0}
\end{figure}

These developments boosted the exploration of \emph{extended} lattice defects on topological platforms, among which grain boundaries (GBs) are the most prominent ones. A GB develops at the interface between two misoriented crystalline grains due to the accumulated elastic stress, and at low angles, it consists of an array of dislocations~\cite{Sutton1995, Han2018}. See Fig.~\ref{fig0}. In turn, by virtue of the hybridization between the localized zero-energy dislocation modes, the GBs can host a wide range of quantum phases in both static~\cite{Slager-PRB2016, amundsen2022} and dynamic~\cite{Salib-2022} settings, when a parent topological insulator (TI) is first-order in nature, featuring gapless modes on the edges or surfaces.

In higher-order topological insulators (HOTIs), accommodating localized states on lower-dimensional boundaries, such as hinges and corners~\cite{BBH-Science2017, BBH-PRB2017, Fang-PRL2017, Langbehn-PRL2017, Schindler-SciAdv2018, Khalaf-PRB2018, Hsu-PRL2018, Matsugatani-PRB2018,  Wang-PRL2019, Trifunovic-PRX2019, Calugaru-2019}, the dislocation modes on the other hand typically move to finite energies, controlled by the relative orientation between its Burgers vector (${\bf b}$) and the axis of inversion (domain wall) of the discrete symmetry breaking Wilson-Dirac (WD) mass, responsible for the higher-order topology~\cite{Roy-Juricic-PRR2021}. Such a nontrivial interplay between the real-space geometry of lattice defects and the momentum-space topology of WD mass propels the current pursuit to unveil its signatures on the emergent electronic bands along GBs in HOTIs.

\begin{figure*}[t!]
\includegraphics[width=0.95\linewidth]{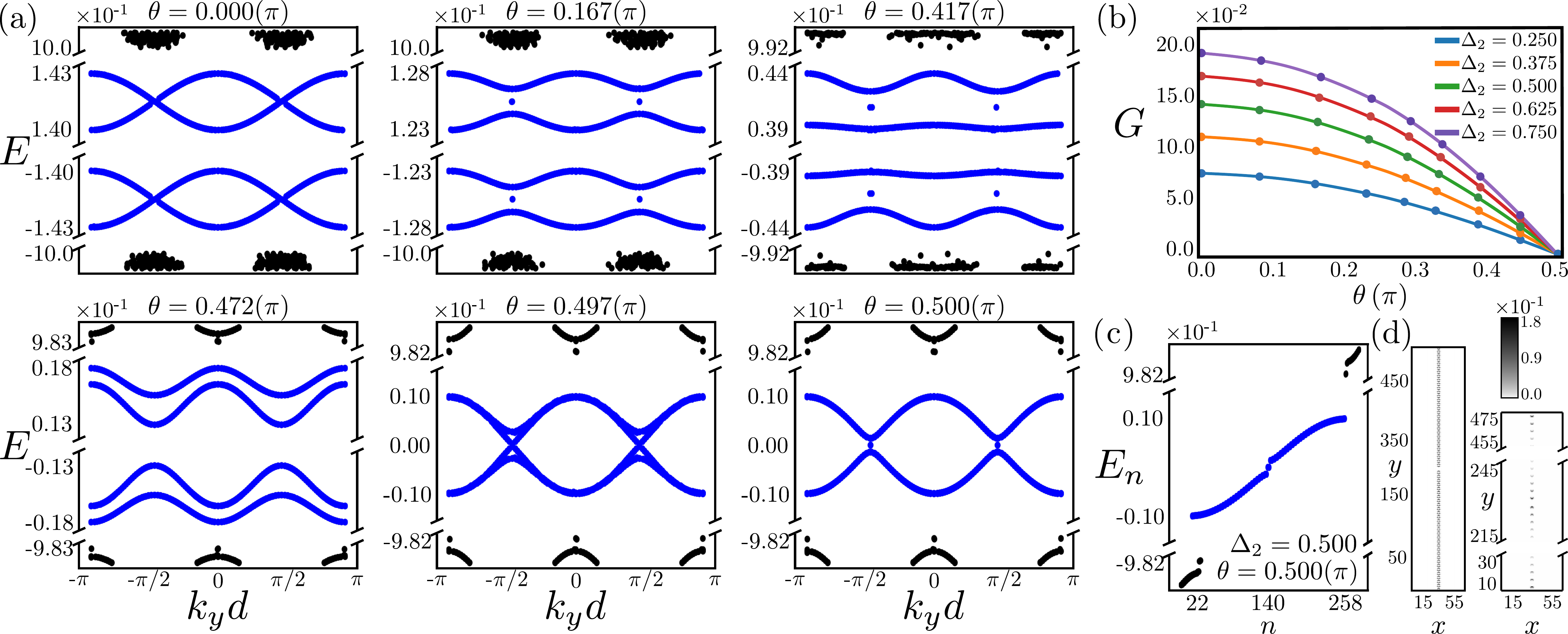}
\caption{Grain boundary (GB) in a 2D second-order topological insulator. (a) Evolution of the GB band structure for the defect (blue) and a few bulk (black) modes as a function of $\theta$ [Eq.~(\ref{eq:WDmass2ndorder})], measuring the relative orientation between the Wilson-Dirac mass domain wall and the dislocation Burgers vector for $\Delta_2=0.5$. (b) Band gap ($G$) between the two closest to zero energy modes, showing that $G \to 0$ as $\theta \to \pi/2$, indicating emergent GB metallicity near the zero energy, or half-filling. (c) Energy spectra showing GB modes (blue) and a few bulk modes (black) at $\theta=\pi/2$. (d) Corresponding local density of states of all the GB modes (left) and two zero-energy modes (right), respectively showing their delocalization along the entire GB and near its ends. Results are obtained for a pair of GB (lower part) and anti-GB (upper part), separated by $10 a$, each containing 30 (anti)dislocations, with periodic boundaries in all directions.
}
\label{fig1}
\end{figure*}

\begin{figure*}[t!]
\includegraphics[width=0.95\linewidth]{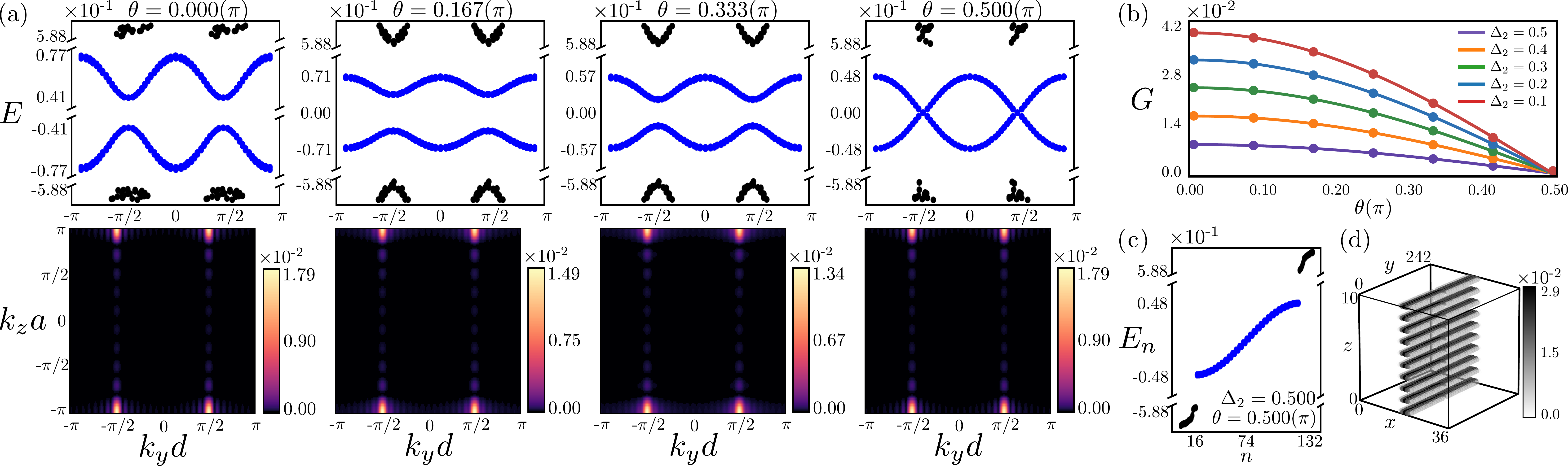}
\caption{Grain boundary (GB) in a 3D second-order topological insulator. (a) Evolution of the GB band structure for the defect (blue) and a few bulk (black) modes as a function of $\theta$ [Eq.~(\ref{eq:WDmass2ndorder})] for $\Delta_2=0.5$, with the spectral weight of the two closest to zero energy modes localized at momenta $(\pm\pi/(2d),\pi/a)$ for any $\theta$. (b) Scaling of the gap ($G$) between two closest to zero energy modes with $\theta$. It vanishes at $\theta=\pi/2$, indicating emergent GB metallicity near the zero energy, or half-filling. (c) Energy spectra showing gapless GB modes (blue) and gapped bulk modes (black). (d) Local density of states for the GB states displaying confinement to the GB plane, on which they are completely delocalized within the 3D bulk, thus yielding a 2D topological metal. Results are obtained for a pair of GB (lower part) and anti-GB (upper part), separated by $10 a$, each containing 15 (anti)dislocations, with periodic boundaries in all directions.}
\label{fig2}
\end{figure*}

\begin{figure*}[t!]
\includegraphics[width=0.95\linewidth]{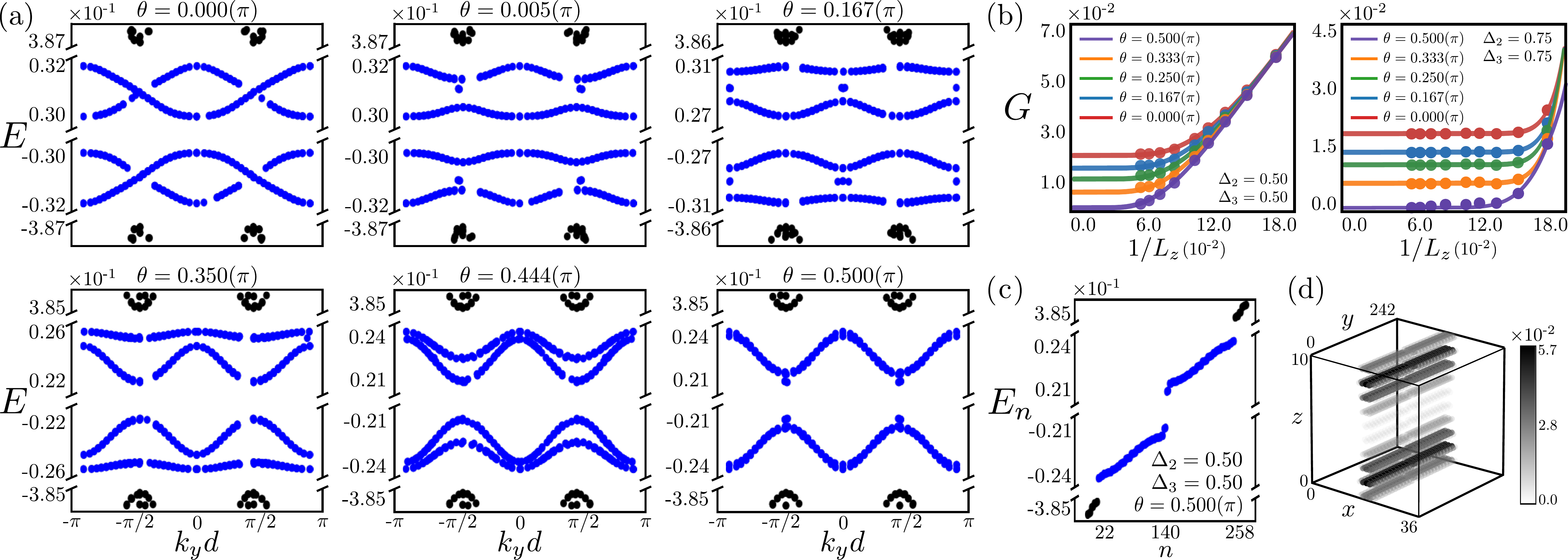}
\caption{Grain boundary (GB) in a 3D third-order topological insulator. (a) Evolution of the band-structure on the top and the bottom edges of the GB plane as the orientation of the second-order mass domain wall moves towards principal crystalline axes $\theta=\pi/2$ [Eq.~\eqref{eq:WDmass2ndorder}] for $\Delta_2=\Delta_3=0.5$. (b) The gap ($G$) between the closest to zero energy states as a function of $\theta$ for two choices of the Wilson-Dirac mass amplitudes ($\Delta_2$ and $\Delta_3$), showing that $G \to 0$ as $\theta \to \pi/2$ (indicating emerging metallicity near the zero energy, or half-filling), but only when the system thickness in the $z$ direction $L_z \to \infty$. For $\theta=\pi/2$, we show the (c) energy spectra supporting the GB modes (blue) and a few bulk modes (black) and (d) local density of states of the GB modes, showing its strong localization near the top and bottom edges of the GB. Results are obtained for a pair of GB (lower part) and anti-GB (upper part), separated by $10 a$, each containing 13 (anti)dislocations, with open (periodic) boundary condition(s) in the $z$ ($x$ and $y$) direction(s).}
\label{fig3}
\end{figure*}

\emph{Key results}.~We show that the GB superlattice in both two-dimensional (2D) and three-dimensional (3D) HOTIs harbors a variety of dispersive bands, which are typically placed at finite energies. However, as the Burgers vector of the constituting edge dislocations and the domain wall directions of the WD mass approach each other these dispersive bands come closer, hybridize and finally touch when the ${\bf b}$-vector points toward gapless corners or hinges. The GB then fosters an emergent self-organized metallicity near the zero energy, stemming from an interplay of lattice geometry and momentum space topology. These outcomes are anchored from the Fourier transformation of the defect modes in the GB mini Brillouin zone (BZ), their local density of states (LDOS) in real space lattice with GB defects and the spectral flow of the defect modes with varying WD mass domain wall in 2D (Fig.~\ref{fig1}) and 3D (Fig.~\ref{fig2}) second-order TIs and 3D third-order TI (Fig.~\ref{fig3}). Our results are thus consequential for the quantum crystals ubiquitously hosting GB defects~\cite{Kim2020}, and metamaterials, where such extended defects can be engineered externally~\cite{Kempkes2019, Imhof2018, dongjuricicroy:2021PRR, Li2018}.

\emph{Lattice model}.~The universal Bloch Hamiltonian for HOTIs in two ($D=2$) and three ($D=3$) dimensions can be split as $\hat{h}_{\rm HOTI} = \hat{h}_{1} + \hat{h}_{\Delta}$. The Hamiltonian for the parent first-order TIs takes the form
\begin{equation}
\hat{h}_{1}= \sum^{D}_{j=1} {\rm d}_j({\bf k}) \Gamma_j + M({\bf k}) \Gamma_{D+1},
\end{equation}
where $d_i({\bf k})=t \sin (k_i a)$, $t$ is the hopping amplitude set to be unity, and $a$ is the lattice spacing. The first-order WD mass, preserving all non-spatial and crystal symmetries, and featuring band inversion (and thus TIs) within the parameter regime $0< \Delta_1/B<4 D$, reads
\begin{align}~\label{eq:firstordermass}
M({\bf k}) = \Delta_1 - 2B \bigg[ D- \sum^{D}_{j=1}\cos(k_j a) \bigg].
\end{align}
A tower of HOTIs can now be constructed by adding discrete symmetry breaking WD masses ($\hat{h}_\Delta$) to $\hat{h}_1$, which can be decomposed as $\hat{h}_\Delta=\hat{h}_{2} + \hat{h}_{3}$. The second-order WD mass in $D=2$ and $D=3$ takes the general form~\cite{Roy-Juricic-PRR2021}
\begin{align}~\label{eq:WDmass2ndorder}
\hat{h}_{2}=\Delta_2  \big\{\cos \theta \; d^{\rm lat}_{x^2-y^2} + \sin \theta  \; d^{\rm lat}_{xy} \big\} \Gamma_{D+2},
\end{align}
where $0 \leq \theta \leq \pi/2$ (about which more in a moment), $d^{\rm lat}_{x^2-y^2}=\cos(k_x a) - \cos(k_y a)$, $d^{\rm lat}_{xy}=\sin(k_x a)\sin(k_y a)$. The third-order WD mass (only in $D \geq 3$) reads as~\cite{NagJuricicRoy2021PRBL}
\begin{align}~\label{eq:WDmass3rdorder}
\hat{h}_{3} = \Delta_3 \left[ 2\cos (k_z a)-\cos (k_x a)-\cos (k_y a) \right] \Gamma_{D+3}.
\end{align}
Here $\Gamma_1, \cdots,\Gamma_{D+2},\Gamma_{D+3}$ are mutually anticommuting Hermitian matrices each of which squares to unity. Results are independent of their explicit representation, which are shown in the Supplementary Information.

The Bloch Hamiltonian $\hat{h}_1 + \hat{h}_2$ describes a second-order TI in both $D=2$ and $3$, since $\hat{h}_{2}$ gaps out the topological edge (surface) states accommodated by the first-order phase, and leaves only four corners (four $z$-directional hinges and two $xy$ surfaces) gapless. As such, for each value of the parameter $\theta$, $\hat{h}_2$ features a domain wall, which lies along the principal axes $k_x=0$ and $k_y=0$ (the diagonals $k_y=\pm k_x$) for $\theta=\pi/2$ ($\theta=0$). Sharp corner or hinge modes then appear only when they lie on the axes of inversion (domain wall) for the WD mass. For additional details on the role of $\theta$, see Ref.~\cite{Roy-Juricic-PRR2021}.

Since $\hat{h}_1 + \hat{h}_2$ involves four (five) mutually anticommuting $\Gamma$ matrices in $D=2$ ($D=3$), their dimensionality is \emph{four}. Consequently, the corner modes are pinned at zero energy due to a unitary (generated by $\Gamma_5$) and an antiunitary particle-hole (PH) symmetry in $D=2$, while in $D=3$ only an antiunitary PH symmetry pins hinge modes to zero energy~\cite{Bitan-PRR2019-1}, as the maximal number of mutually anticommuting four-dimensional Hermitian $\Gamma$ matrices is \emph{five}. Notice that $\hat{h}_1 + \hat{h}_2$ also enjoys the antiunitary composite $C_4 {\mathcal T}$ symmetry, a product of the four-fold rotation about the $z$-axis ($C_4$), generated by $i \Gamma_1 \Gamma_2$ and under which $(k_x,k_y) \to (-k_y,k_x)$, and the time-reversal (${\mathcal T}$) in both $D=2$ and $D=3$. The explicit forms of the antiunitary PH and ${\mathcal T}$ symmetry generators however depend on the $\Gamma$ matrix representation. The model Hamiltonian $\hat{h}_1 + \hat{h}_2$ also breaks the unitary parity (inversion) symmetry (${\mathcal P}$) under which $\vec{k} \to -\vec{k}$. But, its explicit form is dimension and $\Gamma$ matrix representation dependent. Therefore, 2D and 3D second-order TIs additionally preserve composite $C_4 {\mathcal P}$ and ${\mathcal P} {\mathcal T}$ symmetries.

When we turn on the mass term $\hat{h}_3$, the Hamiltonian $\hat{h}_1+\hat{h}_2+\hat{h}_3$ (with finite $\Delta_2$ and $\Delta_3$) describes a third-order topological insulator in $D=3$. Since it requires \emph{six} mutually anticommuting $\Gamma$ matrices, they must therefore be at least eight-dimensional. The additional mass term ($\hat{h}_3$) gaps out otherwise gapless hinge and $xy$ (top and bottom) surface states, yielding the eight zero-energy corner-localized modes in the cubic geometry. For the sharp corner localization, eight corners of the cubic lattice must coincide with the directions along which $\hat{h}_2=0=\hat{h}_3$~\cite{royjuricicoctupole:2021PRBL}. The corner modes in a third-order TI are pinned at zero energy due to a unitary (generated by $\Gamma_7$) as well as an antiunitary PH symmetry. The 3D third-order TI also breaks the individual ${\mathcal T}$, $C_4$, and ${\mathcal P}$ symmetries, and preserves the composite $C_4 {\mathcal T}$, $C_4 {\mathcal P}$, and ${\mathcal P} {\mathcal T}$ symmetries.

Since HOTIs are obtained as descendants of the first-order TIs upon systematically switching on the discrete symmetry breaking WD masses ($\hat{h}_2$ and $\hat{h}_3$), in what follows, we consider the translationally active ${\rm M}$ and ${\rm R}$ phases of the 2D and 3D first-order TIs, respectively, featuring band inversion at finite momentum (${\bf K}_{\rm inv}$) ${\rm M}=(1,1)\pi/a$ and ${\rm R}=(1,1,1)\pi/a$ points in the corresponding BZ of the parent square and cubic lattices. For numerical calculations in $D=2$ and $D=3$, we throughout set $\Delta_1/B=6$ and $\Delta_1/B=11$, respectively [Eq.~(\ref{eq:firstordermass})].

\emph{GB: Construction}.~When introduced in a TI, the elementary building block of a GB defect, a single dislocation, sources an effective hopping phase across the defect $\Phi_{\rm dis}={\bf K}_{\rm inv}\cdot {\bf b}\,\,({\rm mod}\,\, 2\pi)$~\cite{Ran-NatPhys2009, Juricic-PRL2012}. Therefore, electrons encircling the defect can pick up a nontrivial hopping phase $\Phi_{\rm dis}=\pi$ in the ${\rm M}$ (${\rm R}$) phase on the square (cubic) lattice, while $\Phi_{\rm dis}=0$ always in the $\Gamma$ phase as ${\bf K}_{\rm inv}=0$ therein. As a result, in a translationally-active first-order phase, a single dislocation hosts localized  modes at zero energy. By contrast, in HOTIs the additional higher-order WD masses typically gap out the dislocation modes and place them at finite energies, unless the Burgers vector pierces a gapless lower-dimensional boundary, while retaining their topological and symmetry protection~\cite{Roy-Juricic-PRR2021}.

When a GB is immersed in a parent translationally-active first-order TI, the localized zero modes at individual dislocations hybridize, giving rise to an emergent topological metal near the half-filling, or zero energy, along the GB superlattice~\cite{Slager-PRB2016}. The emergence of such a self-organized topological metal can be corroborated from the Fourier transform of the GB modes with respect to the superlattice periodicity $d$, yielding a band structure in the superlattice mini BZ. When the dislocation modes hybridize, they also develop a comparable weight at the middle of two successive defect cores. Thus, we denote the distance between them by $2d$ (Fig.~\ref{fig0}). Next, we report the nature of emergent metallic states in the GB defect consisting of edge dislocations in both 2D (Fig.~\ref{fig1}) and 3D (Fig.~\ref{fig2}) second-order TIs, and 3D third-order TI (Fig.~\ref{fig3}), showcasing an intriguing interplay between the defect geometry and the topology of discrete symmetry breaking WD masses ($\hat{h}_2$ and $\hat{h}_3$). For concreteness, throughout the Burgers vector of constituting dislocation in the GB is taken to be ${\bf b}=a{\bf e}_x$. Here we consider only small angle GBs, characterized by $\phi=\sin^{-1}(a/(2d)) \approx 14.47^\circ$ ($< 15^\circ$) for $d=2a$.

\emph{GB in 2D second-order TI}.~The evolution of the GB band structure as the orientation ($\theta$) of the WD mass domain wall is varied with respect to ${\bf b}$ is shown in Fig.~\ref{fig1}(a). When $\theta=0$, the axes of inversion for the second-order WD mass ($\hat{h}_2$) is \emph{perpendicular} to ${\bf b}$, and a single dislocation features PH symmetric localized modes at finite energies~\cite{Roy-Juricic-PRR2021}. Hybridization among them then yields isolated dispersive bands centered at finite energies along the GB superlattice, separated by a gap $G$ [Fig.~\ref{fig1}(b)]. It measures the energy difference between the bottom and top of such dispersive bands living within the conduction and valence bands, respectively. As $\theta \to \pi/2$, these bands come close to each other and start to hybridize. For $\theta \approx \pi/2$, when the domain wall of $\hat{h}_2$ is \emph{parallel} to ${\bf b}$, this gap eventually closes, yielding an emergent metal around the zero energy. For any $\theta$, the dispersive or metallic states are always separated from the bulk states, like their parent dislocation modes. For $\theta=\pi/2$, besides the dispersive metallic states a pair of zero-energy modes also appear in the spectra [Fig.~\ref{fig1}(a) and (c)]. All of them are \emph{delocalized} along the GB, but confined in its close vicinity, as shown from their LDOS in Fig.~\ref{fig1}(d). The near-zero energy modes, as well as the rest of the GB modes, do not show any spectral weight in the region between the GB and anti-GB, separated by a large distance $10a$, such that the modes bound to them do not overlap with each other.

\emph{GB in 3D second-order TI}.~In $D=3$, a GB consists of an array of edge dislocations, which forms a plane spanned by the directions of the dislocation line and the dislocation array, respectively in the $z$-axis and $y$-axis for ${\bf b}=a {\bf e}_x$. Each dislocation line then hosts localized modes along the $z$ direction, which are typically at finite energies for $\theta \neq \pi/2$, unless ${\bf b}$ points toward gapless hinges ($\theta=\pi/2$), when the dislocation modes become gapless~\cite{Roy-Juricic-PRR2021}. These modes, when hybridize, yield a plethora of 2D dispersive bands along the GB defect, manifesting a gap depending on the parameter $\theta$, as shown in the upper panel of Fig.~\ref{fig2}(a), with the metallic band structure centered around the zero energy emerging for $\theta=\pi/2$. Furthermore, as shown in the lower panel of Fig.~\ref{fig2}(a), the closest to zero-energy modes feature almost the entire spectral weight at momenta $(\pm \pi/(2d),\pi/a)$, which is independent of the parameter $\theta$. The emergent metallicity around the zero energy is further corroborated from the $\theta$-dependence of the gap ($G$) between bottom and top of the dispersive bands respectively residing within the conduction and valence bands, showing that $G$ goes to zero as $\theta \to \pi/2$ [Fig.~\ref{fig2}(b)]. The GB modes are always well separated from the bulk states for any $\theta$ [Fig.~\ref{fig2}(a)], which we also explicitly show for $\theta=\pi/2$ in Fig.~\ref{fig2}(c). Finally, these modes are highly localized in the $yz$-plane constituted by the GB (for any $\theta$), as explicitly shown from their LDOS for $\theta=\pi/2$ in Fig.~\ref{fig2}(d). Appearance of two gapless Dirac points at $(\pm \pi/(2d),\pi/a)$ when $\theta=\pi/2$ in a 2D GB mini BZ conforms to the Nielsen-Ninomiya Fermion doubling theorem~\cite{nielsen-ninomiya}.

\emph{GB in 3D third-order TI}.~An edge dislocation in a 3D third-order TI harbors modes localized near its ends on the top and bottom surfaces, for example, when ${\bf b}=a {\bf e}_x$, which are at finite energies, unless the Burgers vector points toward eight gapless corners, when they become gapless~\cite{Roy-Juricic-PRR2021}. Consequently, a GB in a 3D third-order TI is expected to host gapped dispersive bands localized near the top and the bottom edges of the defect, which become gapless only when $\theta=\pi/2$. Indeed Fig.~\ref{fig3}(a) confirms it, showing that the gap between the dispersive GB bands decreases as the WD mass domain wall approaches the principal crystallographic axes ($\theta \to \pi/2$), when it should become gapless. A \emph{small} residual gap between these dispersive modes at $\theta=\pi/2$ is purely due to a finite thickness of the system in the $z$-direction ($L_z$), which approaches zero as $L_z \to \infty$. See Fig.~\ref{fig3}(b). The GB modes (gapped dispersive or gapless metallic about the half-filling) are always well-separated from the bulk states [Fig.~\ref{fig3}(a)], as explicitly shown in Fig.~\ref{fig3}(c) and localized near the top and bottom surfaces, as shown in Fig.~\ref{fig3}(d) for $\theta=\pi/2$. The fact that these modes are maximally localized just below (above) the top (bottom) surfaces possibly stems from their parent corner modes, also localized slightly away from the terminal surfaces.

\emph{Discussion \& outlook}.~Here we show that 2D and 3D HOTIs foster a variety of dispersive bands within the conduction and valence bands of the emergent BZ constituted by the GB superlattice, with the gap between them tunable by the relative orientation of the WD mass domain wall and the Burgers vector of individual dislocations. Especially, when the Burgers vector pierces gapless corners or hinges, these dispersive bands touch each other at zero energy, giving birth to self-organized topological metals around zero energy confined to the GB defect. See Figs.~\ref{fig1}-\ref{fig3}. The GB modes (dispersive and metallic) are \emph{robust} against weak on-site disorder, the dominant source of elastic scattering in any real material. Explicit results are shown in the Supplementary Information.

Note that the bulk gap is determined by the first-order [Eq.~\eqref{eq:firstordermass}] and higher-order [Eqs.~\eqref{eq:WDmass2ndorder} and~\eqref{eq:WDmass3rdorder}] masses, while the energy scale of the GB modes is set by the latter ones and $\theta$ [Eq.~\eqref{eq:WDmass2ndorder}]. As all the $\Gamma$ matrices appearing in the universal model Hamiltonian for HOTIs mutually anticommute with each other, there exists a finite energy separation between the bulk and GB modes ($\Delta_{\rm bulk-GB}$). However, as $\Delta_2$ and/or $\Delta_3$ become sufficiently large, the higher-order mass overwhelms the first-order mass, and $\Delta_{\rm bulk-GB} \to 0$, but it never vanishes. The explicit dependence of $\Delta_{\rm bulk-GB}$ on the amplitude of higher-order masses is shown in the Supplementary Information.

While here we focus on small angle GBs, described by an array of dislocations (Fig.~\ref{fig0}), at large opening angles ($\phi > 15^\circ$) a GB defect may be described as an array of elementary disclination defects~\cite{Nazarov-PRB2000} or in terms of a lattice of partial dislocations with stacking faults~\cite{King1984}. As a single disclination~\cite{benalcazardiclinationHOTI, Geier-SciPost2021} and a partial dislocation~\cite{Fulga-PRL2019} defect can also harbor localized topological modes, the possibility of emergent metallicity on high-angle GBs stands as a fascinating avenue for future investigation.

The prerequisite for the realization of the proposed emergent metallicity along the GB in HOTIs is a band-inversion at a finite momentum in the BZ (translationally active~\cite{Juricic-PRL2012, slager-natphys2013}) such that ${\bf K}_{\rm inv} \cdot {\bf b}=\pi$ (modulo $2\pi$), as GB defects are rather ubiquitous in quantum crystals. We emphasize that a metallic behavior has been recently observed at GBs in 1T’-MoTe$_2$~\cite{Kim2020}, while some of the realized HOTIs are of translationally-active type~\cite{Schindler2018, Noguchi2021}, which should motivate further investigation of the materials prospects for the realization of our proposal. On the other hand, in metamaterials the predicted GB band structures can be engineered by artificially tuning the tunneling processes and manipulating defects therein. Among them designer metamaterials~\cite{Kempkes2019}, topolectric circuits~\cite{Imhof2018, dongjuricicroy:2021PRR}, photonic~\cite{Li2018} and mechanical~\cite{Grunberg2020} lattices are the most promising platforms. While the metallic nature of the GB modes can be probed via electronic transport measurements in quantum and designer crystals, classical metamaterials can only reveal their dispersive nature from the energy-conserved momentum relation of the associated classical modes, such as the vibrational ones in mechanical lattices. By contrast, the local density of states of the GB modes can be probed in quantum and designer crystals via scanning tunneling spectroscopy, as well as in classical metamaterials, through the measurements of local electric (in topolectric circuits) or mechanical (in mechanical lattice) impedance or two-point pump-probe spectroscopy (in photonic lattices). Our findings should therefore motivate experimental efforts for the realization of the defect-based emergent band structures in a wide range of both quantum and classical topological materials.

\emph{Acknowledgments}.~D.J.S.\ and B.R.\ were supported by NSF CAREER Grant No.\ DMR- 2238679 of B.R. V.J. acknowledges support
of the Swedish Research Council (VR 2019-04735) and Fondecyt (Chile), Grant No.\ 1230933. Nordita is partially supported by Nordforsk.

\emph{Author contributions}.~ D.\ J.\ S.\ performed all the numerical calculations. V.\ J.\ and B.\ R.\ conceived and structured the project, and wrote the manuscript. B.\ R.\ supervised the project.

\emph{Conflict of interests}.~The authors declare no conflicts of interest.

\emph{Data availability}.~The datasets used and/or analysed during the current study available from the corresponding authors on reason- able request. Main codes and the data for generating the figures presented in the main text and Supplementary Information are already available at https://doi.org/10.5281/zenodo.8341312.

\bibliography{references}

\end{document}